\documentclass[%
reprint,
superscriptaddress,
 amsmath,amssymb,
 aps,
]{revtex4-2}

\usepackage{graphicx}
\usepackage{dcolumn}
\usepackage{bm}

\begin{document}

\preprint{APS/123-QED}

\title{Interlayer ferromagnetic order revealed through pressure-enhanced interlayer exchange coupling in Fe$_3$GaTe$_2$}

\author{Riku Iimori}
\email{iimori.riku.es@osaka-u.ac.jp}
\affiliation{Graduate School of Engineering Science, The University of Osaka, 1-3 Machikaneyama, Toyonaka, Osaka 560-8531, Japan}

\author{Takashi Kimura}
\email{kimura.takashi.es@osaka-u.ac.jp}
\affiliation{Graduate School of Engineering Science, The University of Osaka, 1-3 Machikaneyama, Toyonaka, Osaka 560-8531, Japan}

\date{\today}

\begin{abstract}
Van der Waals ferromagnets possess finite interlayer magnetic coupling despite their layered crystal structures, and this coupling plays an important role in determining their magnetic properties. In this study, we performed magnetization measurements on high-quality Fe$_3$GaTe$_2$ single crystals while selectively tuning the interlayer coupling using hydrostatic pressure. We identified two characteristic temperatures, $T_{\rm inf1}$ and $T_{\rm inf2}$, below the Curie temperature with opposite pressure dependences: $T_{\rm inf1}$ is associated with thermally activated domain-wall depinning, whereas $T_{\rm inf2}$ reflects the formation of interlayer ferromagnetic order. Both temperatures exhibit pronounced pressure dependences, in sharp contrast to the nearly pressure-independent Curie temperature associated with intralayer ferromagnetic order. First-principles calculations revealed a substantial pressure-induced enhancement of the interlayer exchange interaction and reasonably reproduced $T_{\rm inf2}$ and its pressure dependence. These results demonstrate that the emergence of interlayer magnetic ordering can be predicted from the microscopic interlayer exchange coupling, establishing a predictive framework for understanding and designing interlayer coupling phenomena in layered magnetic materials.
\end{abstract}

\maketitle


\section{\label{sec:level1}Introduction}
Two-dimensional van der Waals (vdW) ferromagnets have emerged as an important platform for exploring fundamental aspects of quantum magnetism, owing to their well-defined crystal structures and reduced dimensionality~\cite{Discovery2DFM2017, Magnetism2D2018, Gong2DFM2019, 2DFMandHeterostructure2019, ReviewMagnetism2D2021, recentProg2DFM2021, Review2DFMandProximityeffects2020, 2DSpin2020, MagneticGenome2022, 2DFM2022, VDWoptSpin2024}.
In these systems, the interplay between exchange interactions, magnetic anisotropy, and thermal fluctuations leads to rich magnetic phenomena that are absent in conventional three-dimensional magnets~\cite{2DFMPark2026}.
In particular, although interlayer coupling in vdW systems is generally weak, it can have a profound impact on macroscopic quantum states. A representative example highlighting the profound impact of interlayer coupling is twisted bilayer graphene, where a relative twist between otherwise identical layers gives rise to emergent correlated electronic phases, including superconductivity, without modifying the intralayer atomic structure~\cite{TBG2018, TBG22018}. This remarkable phenomenon demonstrates that weak interlayer interactions can fundamentally reconstruct collective quantum states and generate emergent properties far beyond the microscopic energy scale of the interlayer coupling itself. Similar phenomena have been observed in vdW magnets, where the macroscopic magnetic state is governed not only by the dominant intralayer exchange interaction but also by comparatively weak interlayer coupling~\cite{MMoire2021, MMoire2024, 2DFMandHeterostructure2019, vdWgapEng2025}. In bilayer CrI$_3$, for example, the interlayer magnetic configuration can switch between antiferromagnetic and ferromagnetic states through changes in stacking geometry or under external stimuli such as magnetic fields and hydrostatic pressure~\cite{CrI3interlayer2018, CrI3interlayer22018, CrI3interlayer2019, CrI3interlayer22019}. Furthermore, pronounced layer-number-dependent variations of the Curie temperature have been reported in several vdW ferromagnets, including CrTe$_2$~\cite{ThicknessDepTcCrTe22021}, CrSiTe$_3$~\cite{ThicknessDepTcCrSiTe32022}, Cr$_2$Ge$_2$Te$_6$~\cite{Discovery2DFM2017}, Fe$_ n$GeTe$_2$ ($n = 3, 4, 5$)~\cite{ThicknessDepTcFe3GeTe2018, ThicknessDepTcFe3GeTe22018, ThicknessDepTcFe4GeTe2023, ThicknessDepTcFe5GeTe2024}, and Fe$_3$GaTe$_2$~\cite{ThicknessDepTcFe3GaTe2024}, indicating that interlayer magnetic interaction plays a crucial role in stabilizing long-range magnetic order. These observations suggest that even weak interlayer interactions can exert a disproportionate influence on macroscopic magnetic properties. However, experimental studies of the onset temperature for interlayer magnetic ordering and its relationship to the microscopic interlayer exchange interaction remain unexplored.

One possible route toward identifying such characteristic energy scales is provided by anomalous magnetization processes observed in layered vdW ferromagnets. In many vdW magnetic systems, temperature-dependent magnetization exhibits inflection points or other anomalies beyond the monotonic behavior expected from a simple mean-field picture with a single dominant magnetic interaction~\cite{CrI32021, VI32019, VI32021, CrSBr2022, CrTe22025}. Although these anomalies are often discussed in terms of magnetic phase transitions, structural phase transition or magnetic-anisotropy-driven spin reorientation phenomena, their microscopic origin remains unclear.
In particular, Fe-based layered chalcogenides such as Fe$_3$XTe$_2$ (X = Ge, Ga) have been reported to exhibit complex temperature-dependent magnetization, often accompanied by multiple anomalies or inflection points~\cite{AnomalyFe3GeTe22025, AnomalyFe4GeTe22023, AnomalyFe3GaTe22024, AnomalyFe3GaTe220242, AnomalyFe3GaTe22025, DomainPinning2019, DomainPinning2024}. 
These anomalies have often been interpreted as signatures of spin reorientation transitions associated with changes in the effective magnetic anisotropy. However, their microscopic origin remains under debate, especially in systems where competing magnetic interactions and domain-related effects may also play a role.
Although interlayer coupling is generally weaker than the intralayer exchange interaction, growing evidence suggests that it can exert a disproportionate influence on the collective magnetic behavior of layered ferromagnets. 
Therefore, it is important to clarify whether interlayer coupling plays an essential role in the anomalous magnetization features observed in Fe$_3$XTe$_2$.
In this context, external pressure provides a powerful means to selectively tune interlayer coupling, owing to its strong sensitivity to interlayer spacing, while leaving the intralayer interactions relatively unaffected~\cite{Iimori2025Pressure, Kapeghian2024Pressure}. From this perspective, pressure techniques offer a unique opportunity to disentangle the respective roles of interlayer coupling and other competing magnetic interactions.

\begin{figure}[t]
\includegraphics{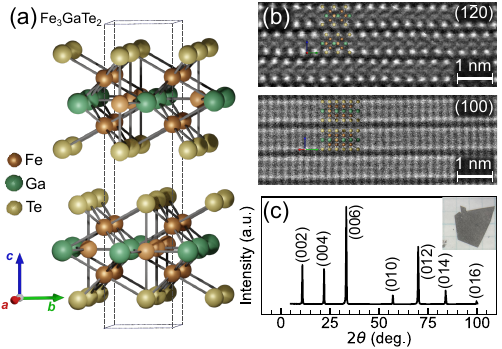}
\caption{\label{fig:1} (a) Schematic illustration of the crystal structure of Fe$_3$GaTe$_2$, showing its layered structure along the crystallographic $c$-axis.
(b) Scanning transmission electron microscopy (STEM) images of the layered structure of Fe$_3$GaTe$_2$ observed on the $(1\bar{2}0)$ and $(100)$ planes.
(c) X-ray diffraction (XRD) pattern of the synthesized single crystal. The inset shows an optical image of the synthesized single crystal.
}
\end{figure}

In this study, we investigate the pressure dependence of magnetization in the vdW ferromagnet Fe$_3$GaTe$_2$, to elucidate the origin of the anomalous temperature-dependent magnetization behavior in terms of interlayer coupling. As a result, we identified two characteristic temperatures, $T_{\rm inf1}\sim160$ K and $T_{\rm inf2}\sim250$ K, which exhibit remarkably different pressure responses: $T_{\rm inf1}$ decreases whereas $T_{\rm inf2}$ increases with increasing pressure, while the Curie temperature remains nearly unchanged. These contrasting behaviors suggest distinct microscopic origins for the two anomalies. First-principles calculations reveal that pressure substantially enhances the interlayer ferromagnetic exchange interaction, whereas the intralayer magnetic interaction is only weakly modified. These results show that calculated exchange interactions reasonably reproduce both the magnitude and pressure dependence of $T_{\rm inf2}$. Our results indicate that $T_{\rm inf1}$ reflects a characteristic temperature for thermally activated domain-wall depinning, whereas $T_{\rm inf2}$ corresponds to the emergence of long-range interlayer ferromagnetic ordering driven by interlayer exchange coupling. These findings establish a quantitative link between microscopic interlayer exchange interactions and macroscopic magnetism in vdW ferromagnets.

\begin{figure*}
\includegraphics{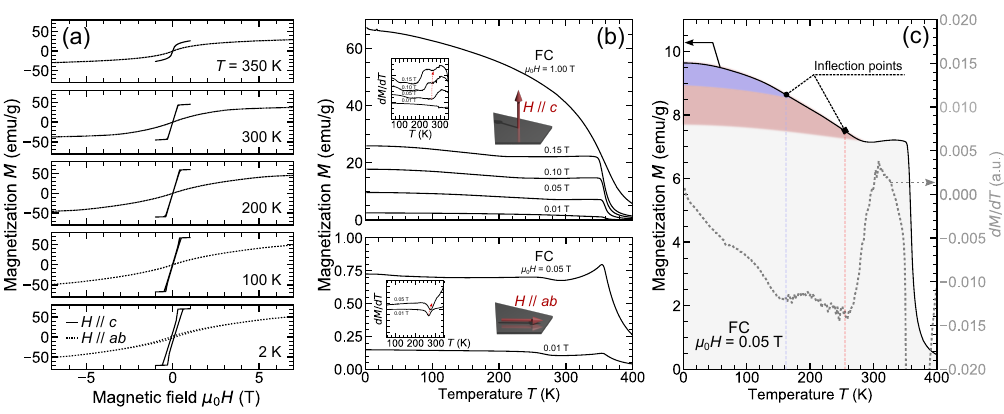}
\caption{\label{fig:2} (a) Magnetization curves measured from 2 K to 350 K. The solid lines represent the magnetization measured under an external magnetic field applied along the crystallographic $c$-axis, while the dashed lines correspond to the in-plane ($ab$-plane) magnetic field configuration.
(b) Temperature dependence of the field-cooled (FC) magnetization measured under various magnetic fields. The magnetization was measured upon warming after field cooling under each applied magnetic field. The upper (lower) panels correspond to measurements with the magnetic field applied parallel to the $c$-axis ($ab$-plane). The inset shows the corresponding derivative curves, $dM/dT$, revealing multiple local minima.
(c) Temperature dependence of the magnetization measured under an external magnetic field of 0.05 T. The derivative signal $dM/dT$ is also shown. Two local minima are observed below the Curie temperature at approximately 160 K and 250 K, suggesting  reconstructions of the magnetic domain properties.
}
\end{figure*}

\section{Experiment}
Fe$_3$GaTe$_2$ single crystals were synthesized by the chemical vapor transport (CVT) method. 
Stoichiometric amounts of Fe, Ga, and Te (3:1:2) were first sealed in an evacuated quartz tube 
($\sim 8 \times 10^{-6}$ Pa) and melted at 1000~$^\circ$C for 24 h to obtain a polycrystalline ingot. 
The ingot was subsequently ground into fine powder.
The resulting powder was then sealed in another evacuated quartz ampoule 
($< 1.0 \times 10^{-4}$ Pa) together with iodine as a transport agent, 
with a concentration of $\sim$ 1.2 mg/cm$^3$. The ampoule was heated at a rate of 
$\sim1~^\circ$C/min and maintained under a temperature gradient, 
with the source side at 770~$^\circ$C and the sink side at 700~$^\circ$C, for 10 days. 
After the growth process, the temperature gradient was removed, and the sample was cooled down to room temperature at a rate of $\sim1~^\circ$C/min.
Experimental details of the crystal synthesis are provided in Sec. S1 of the Supplementary Information.
As a result, plate-like single crystals with a metallic appearance were obtained, as shown in the inset of Fig.~\ref{fig:1}(c). The obtained crystals were stable in air and exhibited typical van der Waals (vdW) characteristics, allowing easy mechanical exfoliation using adhesive tape. 

The atomic structure was investigated by high-angle annular dark-field scanning transmission electron microscopy (HAADF-STEM; JEM-ARM300F, JEOL Ltd.). Specimens for cross-sectional transmission electron microscopy (TEM) observations were prepared by focused ion beam (FIB) milling using a dual-beam FIB-SEM system (Helios 5 UX, Thermo Fisher Scientific). Crystal structural characterization was performed by X-ray diffraction (XRD) using Cu K$\alpha$ radiation (D8 ADVANCE, Bruker).  Magnetic properties were measured using a magnetic property measurement system (MPMS, Quantum Design). Measurements were conducted over a temperature range of 2–400 K under magnetic fields up to 7 T.
Magnetization measurements under pressure were performed using a low-background pressure cell consisting of a CuBe cylinder, a zirconia piston, and a Teflon tube. 
Daphne oil 7373 (Idemitsu Kosan Co., Ltd.) was used as the pressure-transmitting medium. 

To gain insight into the microscopic origin of the pressure-dependent magnetic properties, first-principles calculations were performed. Electronic structure calculations for Fe$_3$GaTe$_2$ were carried out within the framework of density functional theory (DFT) using the QUANTUM ESPRESSO package~\cite{QE1, QE2}. Projector-augmented-wave (PAW) pseudopotentials together with the generalized gradient approximation in the Perdew-Burke-Ernzerhof (PBE) form were employed for the exchange--correlation functional. The plane-wave basis cutoff energy was set to 70 Ry, and a $12 \times 12 \times 4$ $k$-point mesh was adopted for the self-consistent calculations. To properly describe interlayer vdW interactions, the DFT-D3 correction scheme proposed by Grimme was included~\cite{DFT-D, DFT-D2}. Under applied pressure, both lattice parameters and internal atomic coordinates were fully optimized until the residual force acting on each atom became smaller than 0.02 eV/\AA.

\section{Results and Discussion}

Figure~\ref{fig:1}(a) shows the crystal structure of Fe$_3$GaTe$_2$. 
Fe$_3$GaTe$_2$ crystallizes in a hexagonal structure with the space group $P6_3/mmc$ and 
forms a layered structure composed of Te-terminated sheets with fully satisfied covalent bonds, 
resulting in the absence of dangling bonds. 
These layers are weakly coupled via van der Waals (vdW) interactions. 
The single Fe$_3$GaTe$_2$ crystals were synthesized using the chemical vapor transport (CVT) method. As shown in Fig. S1(c) of the Supplementary Information, a large number of plate-like single crystals were successfully obtained. The typical lateral size of the crystals ranges from a few millimeters to approximately 1~cm. Furthermore, the crystals can be easily exfoliated using adhesive tape, reflecting the characteristic layered nature of the material. In addition, characteristic angles of $30^\circ$ and $120^\circ$, originating from the hexagonal crystal symmetry, were clearly observed.
Figure~\ref{fig:1}(b) shows a cross-sectional TEM image of the synthesized Fe$_3$GaTe$_2$ single crystal. The atomic arrangements corresponding to the $(1\bar{2}0)$ and $(100)$ planes were clearly identified, together with a well-defined layered structure.
Figure~\ref{fig:1}(c) shows the X-ray diffraction (XRD) pattern of the obtained single crystal. As can be clearly seen, only the (00$l$) reflections are observed, indicating that the crystal is highly oriented along the $c$-axis and exhibits high crystallinity.
The chemical composition of the synthesized crystals was examined by energy-dispersive x-ray spectroscopy (EDX) analysis, yielding an atomic ratio of ${\rm Fe:Ga:Te} = 45.71:17.98:36.31$, indicating an almost stoichiometric composition of Fe$_3$GaTe$_2$.

\subsection{Magnetic anomalies at ambient pressure}

First, we evaluated the magnetization curves of the synthesized single crystals under ambient pressure.
Figure~\ref{fig:2}(a) shows the magnetic field dependence of the magnetization of the synthesized crystals measured at various temperatures. The solid lines correspond to the magnetic field applied parallel to the $c$-axis, while the dashed lines represent the in-plane field direction.
These results indicate that the synthesized single crystals exhibit strong perpendicular magnetic anisotropy along the $c$-axis. The saturation field along the $c$-axis is approximately 0.6 T, whereas the magnetization does not reach saturation even under an in-plane magnetic field of up to 7 T.

Figure~\ref{fig:2}(c) shows the temperature dependence of the magnetization measured upon warming under an external magnetic field of 0.05~T after field cooling under the same field. First, a sharp change in the magnetization is observed around $T_{\rm C} \sim 350$~K, indicating the ferromagnetic-to-paramagnetic phase transition. In addition to the transition at $T_{\rm C}$, distinct anomalies are also observed at lower temperatures below $T_{\rm C}$. Specifically, local minima in the derivative signal $dM/dT$ are identified at approximately 160~K and 250~K. Since these measurements were performed under a magnetic field much smaller than the saturation field, these anomalies are considered to reflect characteristic changes in the magnetic domain.
In previous studies, similar anomalies below the Curie temperature in the temperature dependence of the magnetization measured under external magnetic fields much smaller than the saturation field have been widely observed in Fe$_3$GaTe$_2$ systems.
Moreover, similar temperature ranges of the inflection-point anomalies have been reported for Fe$_3$GaTe$_2$ crystals synthesized by various growth methods~\cite{ThicknessDepTcFe3GaTe2024, AnomalyFe3GaTe22024, AnomalyFe3GaTe220242, AnomalyFe3GaTe22025, DiscoveryFe3GaTe22022}.
On the other hand, despite the repeated observation of these anomalies in Fe$_3$GaTe$_2$, their microscopic origin has not yet been fully understood.

\begin{figure}[b]
\includegraphics{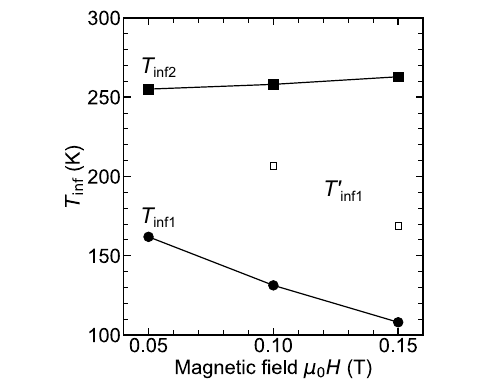}
\caption{\label{fig:3} Magnetic-field dependence of the characteristic temperatures $T_{\mathrm{inf1}}$ and $T_{\mathrm{inf2}}$ observed in Fe$_3$GaTe$_2$.
}
\end{figure}

Figure~\ref{fig:2}(b) shows the temperature dependence of the magnetization measured under various external magnetic fields. At a higher magnetic field of 1~T, where the magnetic moments are nearly saturated along the $c$-axis, the magnetization decreases monotonically toward $T_{\rm C}$ without showing pronounced anomalies. Therefore, these results suggest that the observed inflection points are closely associated with changes in the magnetic domain configuration.
Figure~\ref{fig:3} shows the magnetic-field–temperature phase diagram constructed from the inflection temperatures. Based on their dependence on the magnetic field, the inflection temperatures that decrease with increasing magnetic field are labeled as $T_{\mathrm{inf1}}$ and $T'_{\mathrm{inf1}}$, while the one that increases with increasing magnetic field is labeled as $T_{\mathrm{inf2}}$.
Among them, both $T_{\rm inf1}$ and $T'_{\rm inf1}$ strongly shift toward lower temperatures with increasing magnetic field, indicating that the corresponding domain configuration becomes unstable under the applied magnetic field. Below $T_{\rm inf1}$, the magnetic hysteresis becomes substantially wider, while the remanent magnetization remains finite as shown in Fig.~\ref{fig:2}(a). These results suggest that magnetic domain-wall pinning effects, which are destabilized by the external magnetic field, play an important role in the origin of $T_{\rm inf1}$. It should be noted that, at a very low magnetic field of 0.01~T, $T_{\rm inf1}$ becomes difficult to distinguish clearly because the magnetic domains are not sufficiently aligned by such a weak field. 
In contrast, $T_{\rm inf2}$ appears around 250~K and exhibits a slight increase with increasing magnetic field. Furthermore, the magnetic hysteresis is nearly absent in this temperature region. These results indicate that $T_{\rm inf1}$ and $T_{\rm inf2}$ originate from distinct competing energy scales in the magnetic domain evolution. While $T_{\rm inf1}$ shows a pronounced decrease with increasing magnetic field, consistent with its close relation to magnetic domain dynamics and coercivity, $T_{\rm inf2}$ increases with increasing magnetic field and exhibits negligible hysteresis, suggesting a different microscopic origin that is not directly associated with domain-wall dynamics.

\subsection{Pressure-induced modulation of magnetic properties}

\begin{figure}[t]
\includegraphics{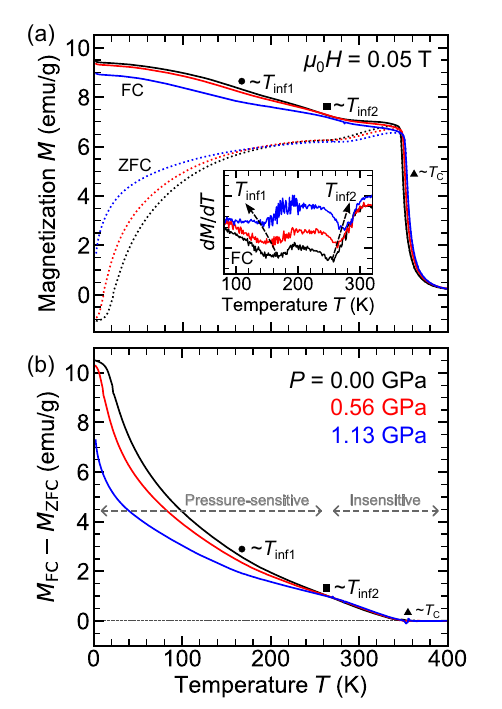}
\caption{\label{fig:4} (a) Temperature dependence of the magnetization measured under pressures from 0.00 to 1.13~GPa. The solid lines represent the field-cooling (FC) process, while the broken lines correspond to the zero-field-cooling (ZFC) process. The applied magnetic field was 0.05~T for all measurements. The inset shows the derivative curves of the magnetization, where the pressure evolution of $T_{\rm inf1}$ and $T_{\rm inf2}$ is indicated by arrows.
(b) Difference between the FC and ZFC magnetization curves under various pressures. The pressure dependence of the magnetization becomes pronounced below $T_{\rm inf2}$, whereas the magnetization remains relatively insensitive to pressure above $T_{\rm inf2}$.
}
\end{figure}

\begin{figure}[t]
\includegraphics{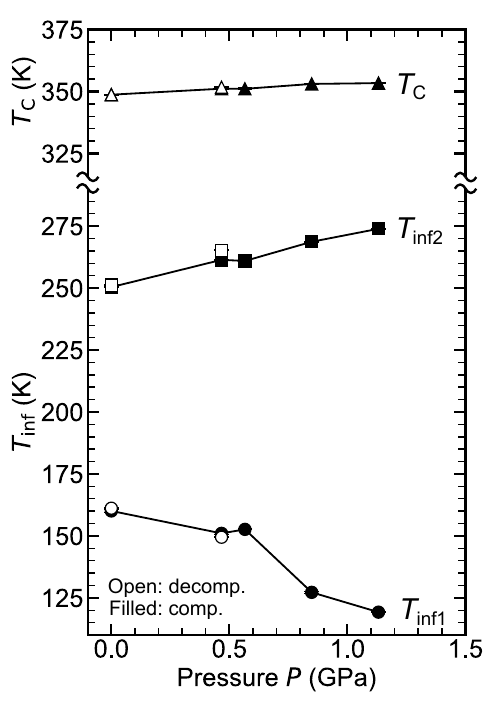}
\caption{\label{fig:5} Pressure dependence of $T_{\rm C}$, $T_{\rm inf1}$, and $T_{\rm inf2}$. Filled markers represent the compression process, while open markers correspond to the decompression process.
}
\end{figure}

\begin{figure}[t]
\includegraphics{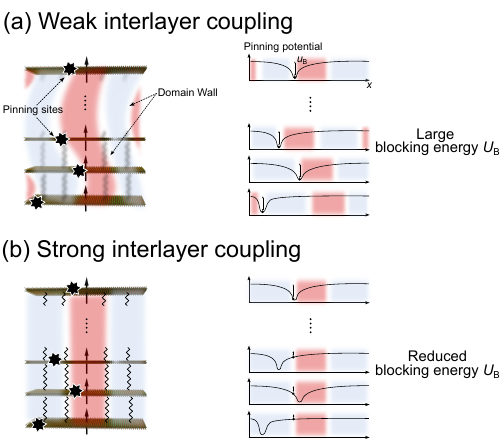}
\caption{\label{fig:6} Schematic illustration of the proposed mechanism for the pressure-induced
decrease in $T_{\mathrm{inf1}}$.
(a) For weak interlayer coupling, domain walls in individual layers can
independently occupy locally favorable pinning sites, resulting in a large
domain-wall blocking energy $U_{\mathrm{B}}$.
(b) Enhanced interlayer coupling correlates the domain-wall positions across
the layers and prevents each domain wall from independently occupying its
locally favorable pinning site, thereby reducing the effective blocking
energy $U_{\mathrm{B}}$.
The reduction in $U_{\mathrm{B}}$ leads to a decrease in
$T_{\mathrm{inf1}}$ within the thermally activated domain-wall relaxation
picture.
}
\end{figure}

\begin{figure*}
\includegraphics{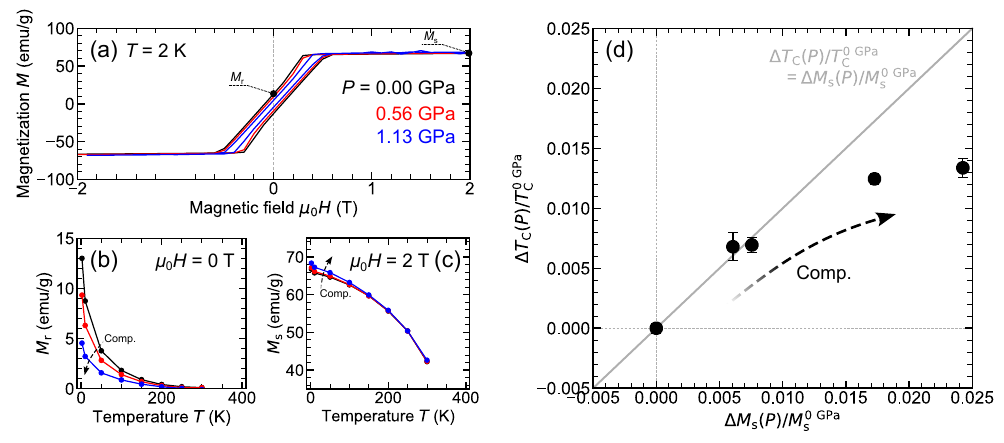}
\caption{\label{fig:7} (a) Magnetization processes under various pressures.
(b) Temperature dependence of the remanent magnetization measured under zero magnetic field for various pressures.
(c) Temperature dependence of the saturation magnetization measured at $\mu_0 H=$ 2~T under various pressures.
(d) Relative changes in the saturation magnetization and the Curie temperature. The solid line represents the relation $\Delta T_{\rm C}(P)/T_{\rm C}^{0~{\rm GPa}} = \Delta M_{\rm s}(P)/M_{\rm s}^{0~{\rm GPa}}$. The experimental data approximately follow this relation in the low-pressure regime, but gradually deviate from it with increasing pressure.
}
\end{figure*}

To further clarify the origin of the two characteristic temperatures, we employed relatively low hydrostatic pressures to selectively tune the interlayer coupling while minimizing changes in the intralayer magnetic interactions. Owing to the weak vdW bonding between adjacent layers, such moderate pressures can effectively modify the interlayer magnetic interaction without substantially perturbing the intralayer magnetism. Figure~\ref{fig:4}(a) shows the temperature dependence of the magnetization measured under various pressures.
Although the Curie temperature of the ferromagnetic phase exhibits only a slight change of a few kelvin under pressure, $T_{\mathrm{inf1}}$ and $T_{\mathrm{inf2}}$ show pronounced pressure dependences; specifically, $T_{\mathrm{inf1}}$ decreases, whereas $T_{\mathrm{inf2}}$ increases with increasing pressure as shown in the inset of Fig.~\ref{fig:4}(a).
The pressure dependences of the inflection points and the Curie temperature observed in this study are consistent with our previous reports on Fe$_3$GaTe$_2$ single crystals grown by the self-flux method, indicating that the observed pressure responses are intrinsic characteristics of Fe$_3$GaTe$_2$~\cite{Iimori2025Pressure}.
The pressure dependence of $T_{\mathrm{inf1}}$ exhibits a trend similar to that observed with increasing applied magnetic field, suggesting that the magnetic domains pinned at low temperatures are destabilized by pressure. In our previous study, we suggested that pressure enhances the interlayer magnetic coupling~\cite{Iimori2025Pressure}. From this perspective, the enhanced interlayer interaction may act as an effective magnetic field, thereby relaxing the pinned domain state.

Figure~\ref{fig:4}(b) shows the temperature dependence of the difference between the field-cooled (FC) and zero-field-cooled (ZFC) magnetization curves, $(M_{\mathrm{FC}}-M_{\mathrm{ZFC}})$. 
In magnetic materials, the FC--ZFC irreversibility is considered to be associated with domain-wall pinning and spin freezing~\cite{FCminusZFC2003}. 
At the Curie temperature, $M_{\mathrm{FC}}-M_{\mathrm{ZFC}}$ approaches zero, indicating the disappearance of magnetic hysteresis and magnetic domains associated with the ferromagnetic state. Below $T_{\rm C}$, in contrast, $M_{\mathrm{FC}}-M_{\mathrm{ZFC}}$ becomes finite, suggesting the formation of magnetic domains. Notably, the pressure response of $M_{\mathrm{FC}}-M_{\mathrm{ZFC}}$ emerges below $T_{\rm inf2}$. Above $T_{\rm inf2}$, $M_{\mathrm{FC}}-M_{\mathrm{ZFC}}$ is relatively insensitive to pressure, whereas a pronounced pressure dependence emerges below $T_{\rm inf2}$. Considering that, in this relatively low-pressure regime, pressure predominantly modifies the interlayer coupling while leaving the intralayer magnetic interactions largely unaffected, this crossover suggests that $T_{\rm inf2}$ marks the onset of interlayer magnetic coherence.

Figure~\ref{fig:5} quantitatively summarizes the pressure dependences of $T_{\mathrm{inf1}}$, $T_{\mathrm{inf2}}$, and $T_{\mathrm{C}}$. As pressure increases, $T_{\mathrm{inf1}}$ decreases by approximately 40 K, whereas $T_{\mathrm{inf2}}$ increases by approximately 25 K, demonstrating remarkably large pressure responses. In contrast, $T_{\mathrm{C}}$, which is primarily governed by the intralayer ferromagnetic interaction, exhibits only a slight increase of approximately 5 K. The pronounced pressure dependence of $T_{\mathrm{inf2}}$ compared with the weak pressure dependence of $T_{\mathrm{C}}$ therefore further supports the interlayer origin of $T_{\mathrm{inf2}}$. These results also indicate that $T_{\mathrm{inf1}}$ is governed by a pressure-sensitive magnetic process distinct from that determining $T_{\mathrm{C}}$. Importantly, these pressure-induced changes reversibly return to their original values upon decompression, indicating that the observed behaviors originate from an intrinsic and reversible modification of the magnetic coupling rather than irreversible structural degradation or sample damage.

\subsubsection{Changes in $T_{\rm inf1}$ and $T_{\rm inf2}$: Interlayer coupling}

We first discuss the pressure dependence of $T_{\mathrm{inf1}}$.
As shown in Fig.~\ref{fig:4}(b), $(M_{\mathrm{FC}}-M_{\mathrm{ZFC}})$
around $T_{\mathrm{inf1}}$ decreases systematically with increasing
pressure. Since $T_{\mathrm{inf1}}$ is closely associated with
domain-wall pinning, its pressure dependence can be understood in terms
of the characteristic domain-wall blocking energy. Assuming a thermally
activated domain-wall relaxation process, the relaxation time can be
expressed as
\begin{equation}
\tau_{\mathrm{DW}}
=
\tau_0
\exp\left(
\frac{U_{\mathrm{B}}}{k_{\mathrm{B}}T}
\right),
\label{eq:DW_relaxation}
\end{equation}
where $\tau_0$ is the characteristic time and $U_{\mathrm{B}}$ is the
effective domain-wall blocking energy determined by the pinning
landscape. At $T_{\mathrm{inf1}}$, the domain-wall relaxation time
becomes comparable to the characteristic experimental time scale
$\tau_{\mathrm{m}}$, i.e.,
$\tau_{\mathrm{DW}}(T_{\mathrm{inf1}})\sim\tau_{\mathrm{m}}$.
Therefore,
\begin{equation}
T_{\mathrm{inf1}}
\simeq
\frac{U_{\mathrm{B}}}
{k_{\mathrm{B}}\ln(\tau_{\mathrm{m}}/\tau_0)}.
\label{eq:Tinf1_barrier}
\end{equation}
Within this thermally activated picture, the experimentally observed
decrease in $T_{\mathrm{inf1}}$ with increasing pressure directly
indicates a reduction in the effective domain-wall blocking energy.
The simultaneous suppression of
$(M_{\mathrm{FC}}-M_{\mathrm{ZFC}})$ is also consistent with the
pressure-induced weakening of domain-wall pinning.
A possible microscopic origin of the reduction in the domain-wall
blocking energy is the pressure-enhanced interlayer magnetic coupling as shown in Fig.~\ref{fig:6}.
In a layered ferromagnet, local defects and magnetic inhomogeneities
generally produce different pinning landscapes in different layers.
When the interlayer magnetic coupling is relatively weak, the domain
wall in each layer can adjust its position relatively independently
to the local pinning landscape. In contrast, enhanced interlayer
coupling favors the alignment of domain-wall positions across
neighboring layers, resulting in a more strongly correlated
domain-wall structure along the $c$-axis. Since the pinning sites are
randomly distributed among the layers, such interlayer correlation
prevents each domain wall from independently occupying its locally
preferred pinning position. Consequently, the layer-dependent random
pinning potentials are effectively averaged over the correlated
domain-wall structure, reducing the effective domain-wall blocking
energy $U_{\mathrm{B}}$. This picture provides a possible microscopic explanation for
the simultaneous decrease in $T_{\mathrm{inf1}}$ and
$(M_{\mathrm{FC}}-M_{\mathrm{ZFC}})$ with increasing pressure.

In contrast, $T_{\mathrm{inf2}}$ exhibits a completely different behavior and increases under pressure. In addition, the pressure dependence of the $(M_{\mathrm{FC}}-M_{\mathrm{ZFC}})$ curves becomes significantly weaker above $T_{\mathrm{inf2}}$. Considering that the dominant effect of pressure in our experimental range of approximately 1 GPa is the enhancement of the interlayer magnetic coupling, these results suggest that $T_{\mathrm{inf2}}$ corresponds to a characteristic temperature above which the influence of the interlayer magnetic coupling on the magnetic domain structure becomes ineffective.
As described above, $T_{\mathrm{inf2}}$ exhibits  increase by the external magnetic field and a remarkably large pressure response. This behavior is consistent with the interpretation that $T_{\mathrm{inf2}}$ represents an upper limit of the temperature range where the interlayer magnetic coupling significantly influences the magnetic domain structure.
In other words, above $T_{\mathrm{inf2}}$, the magnetic domain structure is considered to become more two-dimensional in nature, with relatively weak interlayer coupling and enhanced layer-by-layer magnetic independence.

Here, we also examine the effect of pressure on the magnetic hysteresis ($M$–$H$) curves.
Figure~\ref{fig:7}(a) shows the magnetization hysteresis curves under various pressures at $T=2$ K.
Pressure induces a softening of the magnetic hysteresis behavior. 
As shown in Figure~\ref{fig:7}(b), the remanent magnetization $M_{\mathrm{r}}$ decreases under pressure over the entire measured temperature range. 
These results support the scenario that pressure suppresses magnetic-domain pinning and promotes domain-wall motion in the bulk crystal.

We should note that our previous measurements on Fe$_3$GaTe$_2$ thin films with thicknesses of approximately 145~nm showed that pressure increases the nucleation field $H_{\mathrm{n}}$ at which the magnetization reversal occurs through abrupt domain nucleation~\cite{Iimori2025Pressure}. This apparently different pressure response can be understood by considering the different magnetization-reversal mechanisms in the bulk and thin-film samples. In the bulk crystal, the magnetization process is strongly influenced by domain-wall motion and pinning, whereas in sufficiently thin samples, the finite thickness restricts the development of the bulk-like interlayer domain structure, making domain nucleation more important for magnetization reversal. In this regime, the nucleation field is sensitive to the perpendicular magnetic anisotropy and interlayer exchange interaction, whose enhancement under pressure can increase $H_{\mathrm{n}}$ even when domain-wall pinning is weakened. 

More importantly, both $T_{\mathrm{inf1}}$ and $T_{\mathrm{inf2}}$, which are clearly observed in the bulk crystals, disappear in the approximately 43-nm-thick samples. This thickness dependence provides further evidence that both anomalies are closely associated with interlayer magnetic effects rather than solely with intralayer magnetism. 
In particular, the disappearance of $T_{\mathrm{inf2}}$ is consistent with its assignment to the development of three-dimensional magnetic correlations through interlayer coupling, since the reduced sample thickness may limit the spatial extent of magnetic correlations along the $c$-axis, thereby obscuring the distinct onset of long-range interlayer magnetic order.
The simultaneous disappearance of $T_{\mathrm{inf1}}$ further suggests that the domain-related magnetic irreversibility is also strongly influenced by the interlayer magnetic coupling. Thus, the distinct magnetization processes and the disappearance of both characteristic temperatures in thin films consistently highlight the important role of interlayer magnetic coupling in the bulk magnetic behavior of Fe$_3$GaTe$_2$. A more detailed discussion of the thin-film results is provided in Sec.~S2 of the Supplemental Material.

\subsubsection{Change in $T_{\rm C}$: Intralayer coupling}

In contrast to the pronounced pressure-induced changes in $T_{\mathrm{inf1}}$ and $T_{\mathrm{inf2}}$, which are associated with the interlayer magnetic coupling, $T_{\mathrm{C}}$ exhibits only a weak pressure dependence. To clarify this contrasting behavior, we examine the pressure dependence of the intralayer ferromagnetic properties through the saturation magnetization $M_{\rm s}$.
As shown in Fig.~\ref{fig:7}(c), $M_{\rm s}$ slightly increases with increasing pressure. The value of $M_{\rm s}$ is mainly governed by the intralayer ferromagnetic interaction and is relatively insensitive to the pressure-induced reduction of the interlayer distance. Although a slight enhancement of $M_{\rm s}$ is observed, this behavior is consistent with the small increase in the Curie temperature. Within a simple itinerant ferromagnetic picture, the Curie temperature is expected to scale with the exchange splitting energy, $T_{\mathrm{C}} \propto \Delta_{\mathrm{ex}}$, while the exchange splitting approximately follows $\Delta_{\mathrm{ex}} \propto I M_{\rm s}$, where $I$ is the Stoner parameter. Therefore, within a crude approximation based on the Stoner model, the pressure-induced change in $T_{\mathrm{C}}$ is expected to scale approximately linearly with the change in $M_{\rm s}$.
As shown in Fig.~\ref{fig:7}(d), $T_{\mathrm{C}}$ indeed shows a small increase with pressure, broadly following the pressure dependence of $M_{\rm s}$. At higher pressures, however, the increase in $T_{\mathrm{C}}$ becomes weaker than that in $M_{\rm s}$, which may indicate a slight reduction in the effective Stoner parameter under pressure. Importantly, the pressure-induced changes in both $T_{\mathrm{C}}$ and $M_{\rm s}$ remain within approximately 2.5\%, indicating that the overall intralayer ferromagnetic properties are only weakly modified within our pressure range of approximately 1~GPa. 

In contrast, $T_{\mathrm{inf1}}$ and $T_{\mathrm{inf2}}$ exhibit substantially larger pressure responses. The pronounced pressure-induced changes in $T_{\mathrm{inf1}}$ and $T_{\mathrm{inf2}}$ cannot be explained solely by the small changes in the intralayer ferromagnetic properties, and instead indicate a significant contribution from the pressure-sensitive interlayer magnetic coupling.

\subsection{Microscopic origin of interlayer magnetic coupling}

\begin{figure*}
\includegraphics{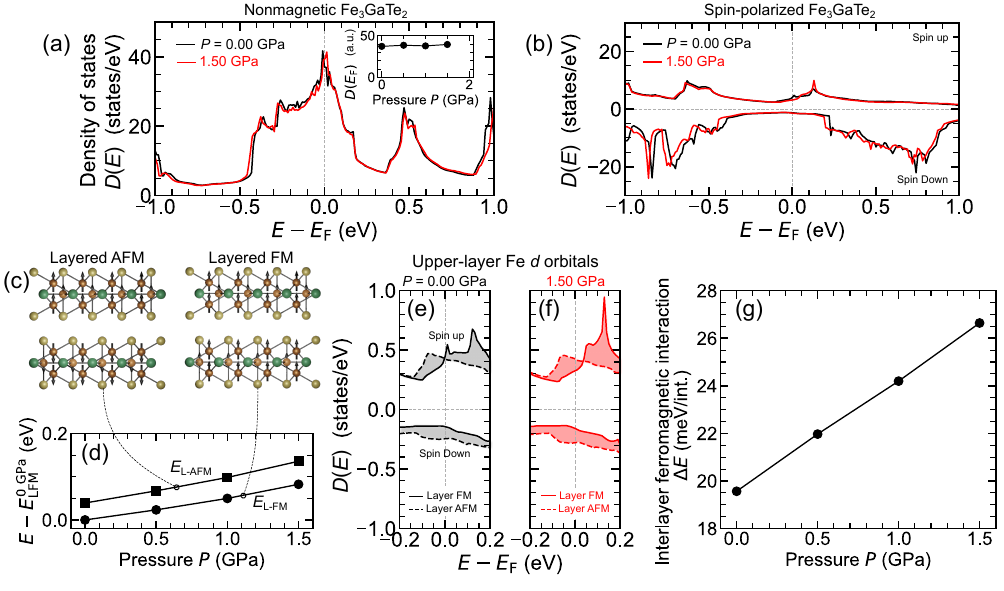}
\caption{\label{fig:8}Electronic structure calculations for understanding the pressure dependence of the magnetic properties.
(a) Density of states (DOS) of non-spin-polarized Fe$_3$GaTe$_2$ calculated at 0.00 and 1.50~GPa. The inset shows the DOS value at the Fermi energy $E_{\rm F}$. Since $D(E_{\rm F})$ does not show significant change up to 1.50~GPa, the pressure-induced change in the Stoner instability, which is responsible for the ferromagnetic instability, is considered to be small.
(b) Spin-polarized density of states in the ferromagnetic state.
(c) Schematic illustrations of the interlayer ferromagnetic and interlayer antiferromagnetic states.
(d) Pressure dependence of the total energies for the interlayer ferromagnetic and interlayer antiferromagnetic states.
(e, f) Density of states of the $d$ orbitals of Fe atoms in the upper layer for the interlayer ferromagnetic and interlayer antiferromagnetic states at 0.00 and 1.50~GPa, respectively. The shaded regions indicate the difference between the two states, reflecting the contribution of the interlayer magnetic interaction.
(g) Pressure dependence of the interlayer ferromagnetic interaction energy.
}
\end{figure*}

To further clarify the microscopic origin of the pressure-induced changes in the interlayer magnetic coupling, first-principles electronic structure calculations were performed.
First, we show that the intralayer ferromagnetic interaction exhibits comparatively weak pressure dependence.
Figure~\ref{fig:8}(a) shows the density of states of nonmagnetic Fe$_3$GaTe$_2$. A pronounced peak structure is observed near the Fermi energy, suggesting stabilization of the ferromagnetic state from the viewpoint of the Stoner instability~\cite{Stoner1938, JungStoner2000, ZhaStoner2023}. On the other hand, no significant change in this instability is observed under pressure up to 1.50~GPa. In fact, the density of states at the Fermi energy $D(E_{\rm F})$ exhibits only weak pressure dependence. This result is consistent with the experimental observation that the intralayer ferromagnetic interaction exhibits only weak pressure dependence. Figure~\ref{fig:8}(b) shows the spin-polarized density of states under pressure. Similar to the nonmagnetic density of states, no substantial pressure-induced modification is observed in the overall electronic structure.

Next, we discuss the interlayer magnetic coupling. Figures~\ref{fig:8}(c) and \ref{fig:8}(d) show the calculated   interlayer ferromagnetic and layered antiferromagnetic states, respectively. The calculations reveal that the layered antiferromagnetic state is energetically less stable than the interlayer ferromagnetic state, indicating that Fe$_3$GaTe$_2$ intrinsically favors interlayer ferromagnetic coupling in its ground state.
The interlayer magnetic coupling energy per interface was evaluated as
$\Delta E=(E_{\mathrm{L\mbox{-}AFM}}-E_{\mathrm{L\mbox{-}FM}})/2$,
where the factor of $1/2$ accounts for the two interlayer interfaces in the unit cell.
The pressure dependence of $\Delta E$ is summarized in Fig.~\ref{fig:8}(g).
As a result, the energy difference between the interlayer ferromagnetic and layered antiferromagnetic states, which is approximately $\Delta E \sim 20$~meV at ambient pressure, monotonically increases by approximately 36\% under pressure, indicating a substantial enhancement of the interlayer ferromagnetic coupling.
To further investigate the microscopic origin of the large pressure-induced enhancement of the interlayer magnetic coupling, we calculated the density of states for both the layered ferromagnetic (L-FM) and layered antiferromagnetic (L-AFM) states. Here, we compare the spin-up density of states in the upper layer shown in Figs.~\ref{fig:8}(e) and \ref{fig:8}(f), namely,
\begin{equation}
\Delta D_{\rm interlayer}(E)
=
D^{\mathrm{upper}}_{\rm{L\mbox{-}AFM}}(E)
-
D^{\mathrm{upper}}_{\rm{L\mbox{-}FM}}(E).
\label{eq:DOS_difference}
\end{equation}
In the absence of interlayer magnetic coupling, the electronic structure of the upper layer is independent of the magnetic configuration of the neighboring layers. In such a case, $\Delta D_{\rm interlayer}(E)=0$.
Therefore, the finite value of $\Delta D_{\rm interlayer}(E)$ directly reflects the influence of the interlayer magnetic coupling on the electronic structure.
The shaded regions in Figs.~\ref{fig:8}(e) and \ref{fig:8}(f) represent $\Delta D_{\rm interlayer}(E)$. Under pressure, $\Delta D_{\rm interlayer}(E)$ near the Fermi energy becomes significantly enhanced. In addition, the shaded regions increases over a wide occupied energy range. These results indicate that pressure enhances the interlayer orbital hybridization and increases the electronic energy gain of the interlayer ferromagnetic state relative to the layered antiferromagnetic state.
Importantly, the enhancement of $\Delta E$ is directly correlated with the pressure-induced increase in the occupied-state spectral redistribution originating from interlayer orbital overlap. This behavior strongly suggests that the interlayer ferromagnetic coupling is governed primarily by short-range exchange interactions mediated by electronic hybridization. In contrast, long-range dipolar interactions are not expected to produce such strong modifications in the electronic density of states or the occupied-state energy distribution. Therefore, the present results indicate that the giant pressure enhancement of the interlayer ferromagnetic coupling originates mainly from pressure-sensitive short-range exchange interactions associated with interlayer orbital overlap, as further supported by the pressure-induced changes in charge- and spin-density distributions in the vdW gap presented in Sec. S4 of the Supplemental Information.

\begin{figure}[t]
\includegraphics{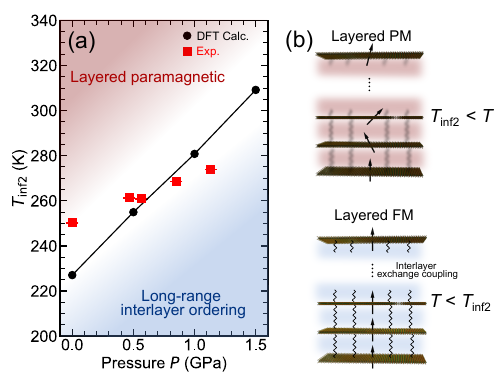}
\caption{\label{fig:9}Interlayer-coupling-induced long-range ferromagnetic ordering in Fe$_3$GaTe$_2$.
(a) Comparison between the experimentally observed and theoretically calculated pressure dependence of $T_{\rm inf2}$.
The theoretical values were obtained from the DFT calculations.
(b) Schematic illustration of interlayer magnetic coherence driven by interlayer magnetic coupling.
Below $T_{\rm inf2}$, long-range interlayer magnetic coherence develops across multiple layers, whereas above $T_{\rm inf2}$, only short-range intralayer magnetic correlations remain.
These results suggest that $T_{\rm inf2}$ characterizes the onset temperature of long-range interlayer magnetic coherence.
}
\end{figure}

One of the most intriguing questions raised by the present results is the origin of the remarkably large pressure-induced shifts of $T_{\rm inf1}$ ($160$~K $\rightarrow$ $119$~K) and $T_{\rm inf2}$ ($250$~K $\rightarrow$ $273$~K). Figure~\ref{fig:9}(a) compares the experimentally observed pressure dependence of $T_{\rm inf2}$ with the characteristic temperature estimated from the DFT-calculated interlayer exchange coupling $\Delta E$, using the relation $k_{\rm B}T_{\rm inf2}\sim\Delta E$. As a result, the calculated values reproduce the experimental trend reasonably well, indicating that the pressure evolution of $T_{\rm inf2}$ is closely linked to the enhancement of the interlayer magnetic coupling. Based on these observations, we propose that $T_{\rm inf2}$ corresponds to the onset temperature of long-range interlayer magnetic coherence. As schematically illustrated in Fig.~\ref{fig:9}(b), below $T_{\rm inf2}$, the interlayer exchange interaction establishes coherent ferromagnetic alignment across multiple layers, whereas above $T_{\rm inf2}$ the magnetic correlations become increasingly layer-independent owing to thermal fluctuations. The substantial increase of $T_{\rm inf2}$ under pressure, together with the large interlayer exchange energy obtained from the DFT calculations, indicates that Fe$_3$GaTe$_2$ possesses strong interlayer magnetic coupling among van der Waals ferromagnets. These results suggest that $T_{\rm inf2}$ characterizes the emergence of long-range interlayer magnetic ordering driven by the interlayer exchange interaction.

In contrast to the increase in $T_{\rm inf2}$ under pressure, $T_{\rm inf1}$ decreases systematically with increasing pressure. These opposite pressure dependences indicate that the enhanced interlayer coupling affects $T_{\rm inf1}$ and $T_{\rm inf2}$ in different ways. As discussed above, the increase in $T_{\rm inf2}$ is associated with the development of interlayer magnetic coherence. In contrast, the decrease in $T_{\rm inf1}$ is attributed to changes in the domain-wall dynamics. Enhanced interlayer coupling correlates the domain walls in adjacent layers, which can reduce the effective energy barrier for domain-wall motion and thereby lower $T_{\rm inf1}$.

\begin{table}[b]
\caption{\label{tab:interlayer}
Comparison of the interlayer exchange coupling energy
$\Delta E$ and its pressure dependence in Fe$_3$GaTe$_2$
and Fe$_3$GeTe$_2$ obtained from our DFT calculations.
}
\begin{ruledtabular}
\begin{tabular}{lcc}
Material &
$\Delta E$ (meV/int.) &
$d\Delta E/dP$ (meV/GPa) \\
\hline
Fe$_3$GaTe$_2$ & 19.6 & 4.72 \\
Fe$_3$GeTe$_2$ & $-1.96$ & $-0.13$ \\
\end{tabular}
\end{ruledtabular}
\end{table}

To further evaluate the significance of the pressure-enhanced interlayer magnetic coupling in Fe$_3$GaTe$_2$, we compare its interlayer exchange energy and pressure response with those of Fe$_3$GeTe$_2$ calculated within the same computational framework, as summarized in Table~\ref{tab:interlayer}. The interlayer exchange energy of Fe$_3$GaTe$_2$ is approximately one order of magnitude larger than that of Fe$_3$GeTe$_2$. 
Here, our calculations yield a negative $\Delta E$ for Fe$_3$GeTe$_2$, indicating an interlayer antiferromagnetic ground state, consistent with several previous studies on Fe$_3$GeTe$_2$~\cite{InterlayerAFMFGeT2017,InterlayerAFMFGeT2021,InterlayerAFMFGeT2023}.
More remarkably, the magnitude of the pressure derivative of the interlayer exchange coupling, $|d\Delta E/dP|$, is enhanced by more than one order of magnitude. These results indicate that the interlayer magnetic interaction in Fe$_3$GaTe$_2$ is not only intrinsically strong but also exceptionally sensitive to variations in the vdW gap.
Such a strong interlayer exchange interaction may also play an important role in determining the high Curie temperature of Fe$_3$GaTe$_2$. Previous studies have shown that monolayer Fe$_3$GaTe$_2$ exhibits a Curie temperature of approximately 240 K, whereas the Curie temperature increases to nearly 350 K in the bulk limit~\cite{Wang2024MonoFGaT}. The substantial enhancement of $T_{\rm C}$ with increasing layer number is consistent with a significant contribution from interlayer magnetic coupling, suggesting that the unusually large interlayer exchange interaction identified in the present work is an important factor in stabilizing high-temperature ferromagnetism in this material.

In this context, the pronounced pressure dependence of $T_{\rm inf2}$ can be naturally understood as a consequence of the strong sensitivity of the interlayer magnetic coupling to the vdW gap. Interestingly, pressure-sensitive magnetic anomalies below $T_{\rm C}$ have also been reported in several layered magnetic materials, including VI$_3$~\cite{VI32019, VI32021}. Although these anomalies have often been discussed in terms of structural distortions, spin reorientation, or complex magnetic phase evolution, the present results suggest that pressure-induced modulation of interlayer magnetic coupling may provide an additional and potentially universal mechanism for generating characteristic temperatures below $T_{\rm C}$ in layered magnetic systems. Further investigations of the relationship between interlayer exchange interactions and low-temperature magnetic anomalies in other vdW magnets will therefore be highly desirable.

Finally, as demonstrated throughout this study, the interlayer exchange coupling exerts a profound influence on the magnetic properties of two-dimensional magnets. This finding establishes vdW-gap engineering as a powerful route for controlling the electronic structure, interlayer magnetic coherence, and emergent dimensionality of two-dimensional magnetic systems. The magnetic anomalies identified in the present work provide a direct and quantitative manifestation of the role of interlayer coupling in determining the magnetic behavior of Fe$_3$GaTe$_2$.
We expect that the methodology presented here will provide a useful platform for investigating the microscopic origin of interlayer magnetism in a wide range of van der Waals magnetic systems.

\section{Conclusions}
In conclusion, we have investigated the pressure dependence of magnetization in the layered ferromagnet Fe$_3$GaTe$_2$ and clarified the distinct microscopic origins of two anomalous temperatures, $T_{\rm inf1}$ and $T_{\rm inf2}$. While $T_{\rm inf1}$ decreases markedly under pressure, $T_{\rm inf2}$ exhibits a substantial positive pressure dependence despite the nearly pressure-independent Curie temperature. First-principles calculations reveal that pressure selectively enhances the interlayer ferromagnetic exchange interaction, whereas the intralayer magnetic interaction remains nearly unchanged. Furthermore, the pressure dependence of the interlayer exchange coupling obtained from DFT calculations is in good agreement with the experimentally observed evolution of $T_{\rm inf2}$.

These results indicate that $T_{\rm inf1}$ originates from pressure-dependent magnetic domain dynamics, whereas $T_{\rm inf2}$ corresponds to the onset of long-range interlayer magnetic ordering driven by interlayer exchange coupling. The substantial pressure response of $T_{\rm inf2}$, together with the large interlayer exchange energy obtained from first-principles calculations, demonstrates that interlayer magnetic interactions play a central role in determining the magnetic properties of Fe$_3$GaTe$_2$.

The present work establishes a direct connection between microscopic interlayer exchange interactions and macroscopic magnetic ordering in a van der Waals ferromagnet. More importantly, our results highlight the importance of interlayer magnetic coupling in controlling collective magnetic phenomena in layered materials. These findings identify vdW-gap engineering as a promising strategy for tuning magnetic ordering, electronic states, and dimensionality in two-dimensional magnetic systems.

\begin{acknowledgments}
We are grateful to Akihiro Mitsuda for his support with the high-pressure experiments.
This work was supported by the New Energy and Industrial Technology Development Organization (NEDO) (Uncharted Territory Challenge 24000474),
the Japan Society for the Promotion of Science (JSPS) KAKENHI Grant Numbers 21H05021, 23KJ1701, 25K23549,
the Japan Science and Technology Agency (JST) (CREST: JPMJCR18J1, SICORP: 22480474),
and the Casio Science Promotion Fundation (J43-39) funds.
\end{acknowledgments}

\nocite{*}


\end{document}